\documentclass[12pt]{article}
\usepackage{amssymb,amsmath}
\begin{document}
\begin{flushright}
\vspace{ 1mm } hep-th/0304174
\\
FIAN/TD/04/08\\
\end{flushright}
\vspace{10mm}
\begin{center}

{\Large \bf Scattering and   pair production by a potential barrier}\\

 \vspace{10mm} A.I.Nikishov ${}^\dag$\\

\vspace{3mm}{$\dag$\it Tamm Department of Theoretical Physics,
              Lebedev Physical Institute,\\
 Leninsky Prospect 53, 119991, Moscow, Russia\\
             e-mail: nikishov@lpi.ru}

    \vspace{2mm}

    \end{center}

\begin{abstract}
Scattering and electron-positron pair production by a one-dimensional
  potential is considered in the framework of the $S-$matrix formalism. 
The solutions of the Dirac equation are classified
 according to frequency sign. 
The Bogoliubov transformation relating the  in- and out-states are given.
We show that the norm of a solution of  the wave equation is determined by
the largest amplitude of its asymptotic form when $x_3\to \pm\infty$.
 For a number of potentials we give the explicit expressions
for the complete in- and out-sets of orthonormalized wave functions. We 
note that in principle virtual vacuum processes in external field
influence the phase of wave function of scattered particle.
\end{abstract}
\section{Introduction}
Scattering and pair production by an external field can be treated ether
by Feynman propagator method [1-2] or in the framework of $S-$ matrix
formalism [3-5]. In the latter method the Bogoliubov coefficients, 
 relating the in- and out-states, determine the
probabilities of all processes in an external field. So, at first we have 
to determine the in- and out- states. Surprisingly enough there is a
controversy on how to choose these states among the stationary solutions
of a wave equation with one-dimensional static potential $A_{\mu}(x_3)$. 
The choice made in [6] and accepted in [7-8] disagrees with that one made 
in [4, 9-11].
For detailed justification of our choice see [9-11].

In this paper we give the in- and out-sets of solutions of the Dirac equation 
with barrier potential and the Bogoliubov transformations.  
 We express the normalization of a stationary  one-dimensional solution via
 the largest amplitude of its asymptotic form when $x_3\to\pm\infty$.
For Sauter type  potential and step potential we give the explicit solutions
of the  Dirac equation and also Bogoliubov coefficients. In contrast to [6-7] 
we do not
 assume that the transverse (to the field) momentum of electron is zero.
 The treatment of spinor case in this paper is similar to that of scalar case 
 in [9].

\section{The choice of in- and out-states}

We consider the one-dimensional potential $A^0(x_3)=-aF(kx_3)$ and assume  
  that the corresponding electrical field $E_3=-\frac{\partial A^0}
{\partial x_3}=akF'(kx_3)$ disappears when $x_3\to\pm\infty$. We use the metric
$\eta_{\mu\nu}={\rm diag(-1,1,1,1)}$
and introduce the kinetic energy $\pi^0(x_3)$ and kinetic momentum
$\pi_3(x_3)$ of a classical particle defined by the expressions
$$
\pi^0(x_3)=p^0-eA^0(x_3),\quad \pi_3(x_3)=\sqrt{\pi^2_0(x_3)-m^2_{\perp}},
\quad m^2_{\perp}=m^2+p^2_1+p^2_2. \eqno(1)
$$
  We also use the notation
$$
\left.\pi^0(x_3)\right|_{x_3\to\pm\infty}=\pi^0(\pm), \quad
\left.\pi_3(x_3)\right|_{x_3\to\pm\infty}=\pi_3(\pm)=\sqrt{\pi_0^2(\pm)-
m_{\perp}^2},\quad\pi^{\pm}(x_3)= \pi^{0}(x_3)\pm\pi_3(x_3). \eqno(2)
$$
 The electron charge is denoted as $e=-|e|$.
We assume for definiteness that $E_3>0$ and consider  two regions:
 electron scattering ($\pi^0(\pm)>m_{\perp}$) and Klein region
$$
\pi^0(-)>m_{\perp}, \qquad \pi^0(+)<-m_{\perp},\eqno(3)
$$
In the Klein region the large positive $x_3$ are accessible only to positrons.

For brevity reasons we often write only the wave function factor depending on
$x_3$ (i.e. we drop the factor $\exp\{i[p_1x_1+p_2x_2-p^0t]\}$).
We denote solutions of the Dirac equation as $f_n(x_3)$, where $(n=p,r)$ and
 $p=(p^0,p_1,p_2,0)$ are the eigenvalues of operators $i\partial_0, 
 -i\partial_1, -i\partial_2$ ; $r=1,2$ indicate spin state.
  $f_n$ can be expressed
 via solutions $Q_p(x_3,\pm\lambda)$ of the squared Dirac equation
 $$
 [\Pi^2+m^2\pm ieE_3(x_3)]Q_p(x_3,\pm\lambda)e^{ipx}=0.       \eqno(4)
 $$
 Here
 $$
 \Pi_{\mu}=-i\partial_{\mu}-eA_{\mu},\quad \vec A=0,
 \quad \lambda=\frac{ea}k
 $$
 and $\pm1$ are the eigenvalues of $\sigma_3$. It follows from (4) that
 ($\varphi=kx_3$)
 $$
[\frac{d^2}{d\varphi^2}+\frac{2eap^0}{k^2}F(\varphi)+\lambda^2F^2(\varphi)+
\frac{p_0^2-m_{\perp}^2}{k^2}-i\lambda F'(\varphi)]Q_p(x_3,\lambda)=0.  \eqno(5)
$$

 We consider solutions with the following boundary conditions
$$
\left.{}_{\pm}Q_{p}\right|_{x_3\to-\infty}=
\exp[\pm i\pi_3(-)x_3],                                     
$$ 
$$
\left.{}^{\pm}Q_{p}\right|_{x_3\to\infty}=
\exp[\pm i\pi_3(+)x_3]                                      \eqno(6)
$$
independently of the sign in front of $\lambda$.
The corresponding $f_n$ solutions are [12]
$$
{}_\pm f_n=[4|\pi_3(-)\pi^{\mp}(-)|]^{-\frac12}[u_r\;{}_{\pm}Q_p(x_3,\lambda)
+\pi^{\mp}(-)u'_r\;{}_{\pm}Q_p(x_3,-\lambda)],
$$
$$
{}^\pm f_n=[4|\pi_3(+)\pi^{\mp}(+)|]^{-\frac12}[u_r\;{}^{\pm}Q_p(x_3,\lambda)
+\pi^{\mp}(+)u'_r\;{}^{\pm}Q_p(x_3,-\lambda)].               \eqno(7)
$$
In the standard representation of $\gamma$-matrixes we have
  $$
     u_1= \begin{bmatrix}
p_1-ip_2\\
m\\
p_1-ip_2\\
-m
    \end{bmatrix},\quad
     u_2=\begin{bmatrix}  
m\\
-p_1-ip_2\\
m\\
p_1+ip_2
    \end{bmatrix},                   
  $$
  $$
  u'_1= \begin{bmatrix}
0\\
1\\
0\\
1
    \end{bmatrix},\quad 
    u'_2=\begin{bmatrix}
1\\
0\\
-1\\
0
    \end{bmatrix}.                   \eqno(8)
$$
All these spinors are orthogonal:
$$
u_1^*u_2=u_1^*u'_2=u_1^*u'_1=\cdots=0,
$$
so that $f_{rp}^*f_{r'p}\propto\delta_{r'r}$. This means that states with
$r\ne r'$ are orthogonal.
The normalization  factors  in (7) are chosen in such a way that the 
 current density along the third axis is equal to unity up to a sign.
 The 4-vector transition current is conserved. For $p^0=p'{}^0$ this means
  that 3-current is independent of $x_3$ and can be evaluated using 
 the asymptotic forms of $f_n$. It is easy to find that
$$
j_3(f_{n'},f_n)=f_{n'}^*\alpha_3f_n,\quad
 j_3({}^{\varepsilon}f_{r,p},{}^{\varepsilon'}f_{r',p})=
 \delta_{\varepsilon\varepsilon'}\delta_{rr'}\left\{\begin{array}{cc}
\varepsilon1,\quad \pi^0(+)>m_{\perp},\\
-\varepsilon1,\quad \pi^0(+)<-m_{\perp},
\end{array}\right. \quad \varepsilon,\varepsilon'=\pm;            \eqno(9)
$$
$$
 j_3({}_{\varepsilon}f_{r,p},{}_{\varepsilon'}f_{r',p})=
 \delta_{\varepsilon\varepsilon'}\delta_{rr'}\left\{\begin{array}{cc}
\varepsilon1,\quad \pi^0(-)>m_{\perp},\\
-\varepsilon1,\quad \pi^0(-)<-m_{\perp},
\end{array}\right..                                           \eqno(10)
$$

To obtain the relation between ${}_{\pm}f_n$  and ${}^{\pm}f_n$, we first
write
$$
{}_+Q_p(x_3,\lambda)=a(\lambda)\;{}^+Q_p(x_3,\lambda)+
b(\lambda)\;{}^-Q_p(x_3,\lambda).                               \eqno(11)
$$
$a(\lambda)$ and $b(\lambda)$ are defined by this equation.
From here by complex conjugation and substitutions $p\to-p$, $e\to-e$, we
get
$$
{}_-Q_p(x_3,\lambda)=a^*(-\lambda)\;{}^-Q_p(x_3,\lambda)+
b^*(-\lambda)\;{}^+Q_p(x_3,\lambda).                            \eqno(12)
$$

Now it can be shown [12] that
$$
{}_+f_n=c'_{1n}\,{}^+f_n+c'_{2n}\,{}^-f_n,
$$
$$
{}_-f_n=\pm c'{}^*_{2n}\,{}^+f_n\pm c'{}^*_{1n}\,{}^-f_n,
\quad \pm|c'_{1n}|^2\mp|c'_{2n}|^2=1,                          \eqno(13)
$$
or equivalently 
$$
{}^+f_n=\pm c'{}^*_{1n}\,{}_+f_n-c'_{2n}\,{}_-f_n,
$$
$$
{}^-f_n=\mp c'{}^*_{2n}\,{}_+f_n+ c'{}_{1n}\,{}_-f_n. \eqno(14)
$$
Here
$$
c'_{1n}=\left(\frac{\pi_3(+)|\pi^{-}(+)|}{\pi_3(-)|\pi^-(-)|}
\right)^{1/2}a(\lambda),\quad
c'_{2n}=\left(\frac{\pi_3(+)|\pi^{+}(+)|}{\pi_3(-)|\pi^-(-)|}\right)
^{1/2}b(\lambda)                                                \eqno(15)
$$
 are independent of spin states $r=1,2$.
The upper sign in front of $c'_{1n}, c'_{2n}$ (and their complex conjugates)
corresponds to scattering region ($\pi^0(\pm)>m_{\perp}$ or
 $\pi^0(\pm)<-m_{\perp}$), while the lower sign corresponds to the Klein
 region ($\pi^0(-)>m_{\perp}, \pi^0(+)<-m_{\perp})$.

The consistency of equations (13-14) can be checked by calculating
$j_3({}_+f_n,{}^+f_n)$ in two different ways: 
$$
j_3({}_+f_n,{}^+f_n)=\pm c'{}^*_{1n}j_3({}_+f_n,{}_+f_n)=
c'{}^*_{1n}j_3({}^+f_n,{}^+f_n).                            \eqno(16)
$$
In the first equation here
the use has been made of the first equation in (14) and equation (10). 
The last term in (16) was obtained  similarly. The last equality  in (16)
is consistent (valid) due to (9).

Now we have to classify the solutions as in- and out-states. For the Klein
 region we have [4], [9]
$$
{}^-\psi_n\equiv{}^-\psi_{n\:out}={\cal N}_n{}_+f_n,
\qquad{}^+\psi_n\equiv{}^+\psi_{n\:out}={\cal N}_n\frac{c'_{2n}{}^*}{c'_{2n}{}
}{}^+f_n,
$$
$$
{}_-\psi_n\equiv{}_-\psi_{n\:in}={\cal N}_n{}_-f_n,
\quad{}_+\psi_n\equiv{}_+\psi_{n\:in}=
{\cal N}_n{}^-f_n.                                        \eqno(17)
$$
Here the subscripts and superscripts $\pm$  in front of
 $\psi$-functions indicate the sign of frequencies i.e. the sign of $\pi^0$
of the largest wave.  ${}^-\psi_n$ is the out-wave because in
 this state only one current goes out of the barrier (two other currents 
 go to the barrier). Similar arguments hold for other states.
The normalization factor
 ${\cal N}_n$ will be determined later, see eq. (40) and text below it.
In terms of these $\psi-$functions in (17) the relations (13) and (14) take on 
the form required for the application in the $S-$matrix theory, see [3-4],
$$
{}_+\psi_n=c_{1n}\,{}^+\psi_n+c_{2n}\,{}^-\psi_n,
$$
$$
{}_-\psi_n=-c^*_{2n}\,{}^+\psi_n+c^*_{1n}\,{}^-\psi_n; 
$$
$$
{}^+\psi_n=c^*_{1n}\,{}_+\psi_n-c_{2n}\,{}_-\psi_n,
$$
$$
{}^-\psi_n=c^*_{2n}\,{}_+\psi_n+c_{1n}\,{}_-\psi_n;  \eqno(18)
$$
$$
|c_{1n}|^2+|c_{2n}|^2=1,\qquad c_{1n}=-\frac{c'_{1n}}{c'{}^*_{2n}},\qquad
c_{2n}=\frac1{c'_{2n}}.                                           \eqno(19)
$$
The sign of frequency is not conserved in the Klein region. We note also 
that the first eq. in (19) follows from the last eq. in (13) with the lower 
signs.

The corresponding Bogoliubov transformation for the creation operators
$a_n^{\dag}$ ($b_n^{\dag}$) and distraction operators $a_n$ ($b_n$) for
particle (antiparticle) are obtainable from the definition of the field
operator
$$
\Psi=\sum_n(a_{n\:in}\,{}_+\psi_n+b^{\dag}_{n\:in}\,{}_-\psi_n)=
\sum_n(a_{n\:out}\,{}^+\psi_n+b^{\dag}_{n\:out}\,{}^-\psi_n).   \eqno(20)
$$
It follows from here and (18-19) that
$$
a_{n\;out}=c_{1n}a_{n\;in}-c_{2n}^*b_{n\;in}^{\dag},
$$
$$
b_{n\;out}^{\dag}=c_{2n}a_{n\;in}+c_{1n}^*b_{n\;in}^{\dag},    \eqno(21)
$$
or
$$
a_{n\;in}=c^*_{1n}a_{n\;out}+c^*_{2n}b_{n\;out}^{\dag},
$$
$$
b_{n\;in}^{\dag}=-c_{2n}a_{n\;out}+c_{1n}b_{n\;out}^{\dag}.     \eqno(22)
$$
  
In the scattering region $\pi^0(\pm)>m_{\perp}$  the sign of frequency 
is conserved, but the sign of kinetic momentum is not. We define 
$$
 {}^+\psi_n(x_3|+)=\frac{c'{}^*_{1n}}{c'_{1n}}N_n{}_+f_n,\quad
  {}^+\psi_n(x_3|-)=N_n{}^-f_n,
  $$
  $$
 {}_+\psi_n(x_3|+)=N_n{}^+f_n,\quad {}_+\psi_n(x_3|-)=N_n{}_-f_n. \eqno(23)
 $$
  (The sign in parentheses coincides with the subscript or
 superscript of the corresponding $f$-function.) ${}^+\psi_n(x_3|\pm)$ are 
 functions with two ingoing waves in the past and one outgoing wave in
 the future. Similarly, ${}_+\psi_n(x_3|\pm)$ are functions with one ingoing 
 wave in the past and two  outgoing waves in the future.
In terms of these $\psi-$function the relations (13-14)
 become
 $$
 {}_+\psi_n(x_3|+)=e_{1n}{}^+\psi_n(x_3|+)+e_{2n} {}^+\psi_n(x_3|-),
 $$
 $$
 {}_+\psi_n(x_3|-)=-e^*_{2n}{}^+\psi_n(x_3|+)+e^*_{1n} {}^+\psi_n(x_3|-),
 $$
 $$
 {}^+\psi_n(x_3|+)=e^*_{1n}{}_+\psi_n(x_3|+)-e_{2n} {}_+\psi_n(x_3|-),
 $$
 $$
 {}^+\psi_n(x_3|-)=e^*_{2n}{}_+\psi_n(x_3|+)+e_{1n} {}_+\psi_n(x_3|-);\eqno(24)
 $$
  $$
e_{1n}=\frac1{c'{}^*_{1n}},\quad e_{2n}=-\frac{c'_{2n}}{c'_{1n,}}, \quad
|e_{1n}|^2+|e_{2n}|^2=1.                                         \eqno(25)
  $$
 The last eq. in (25) is equivalent to the last eq. in (13) with upper signs.
Proceeding as usual, from these relations and definition (see eq.(20))
$$
a_{n\:in}(+)\,{}_+\psi_n(x_3|+)+a_{n\:in}(-)\,{}_+\psi_n(x_3|-)=
a_{n\:out}(+)\,{}^+\psi_n(x_3|+)+a_{n\:out}(-)\,{}^+\psi_n(x_3|-),   \eqno(26)
$$
we get
$$
a_{n\:out}(+)=e_{1n}a_{n\:in}(+)-e_{2n}^*a_{n\:in}(-),
$$
$$
a_{n\:out}(-)=e_{2n}a_{n\:in}(+)+e_{1n}^*a_{n\:in}(-),  \eqno(27)
$$
and from here
$$
a^{\dag}_{n\:out}(+)=e^*_{1n}a^{\dag}_{n\:in}(+)-e_{2n}a^{\dag}_{n\:in}(-),
$$
$$
a^{\dag}_{n\:out}(-)=e^*_{2n}a^{\dag}_{n\:in}(+)+e_{1n}a^{\dag}_{n\:in}(-).
                                                                 \eqno(28)
$$
Solving (27) for $a_{n\;in}(\pm)$, we obtain
$$
a_{n\:in}(+)=e_{1n}^*a_{n\:out}(+)+e_{2n}^*a_{n\:out}(-),
$$
$$
a_{n\:in}(-)=-e_{2n}a_{n\:out}(+)+e_{1n}a_{n\:out}(-),  \eqno(29)
$$
and for the creation operators
$$
a_{n\:in}^{\dag}(+)=e_{1n}a_{n\:out}^{\dag}(+)+e_{2n}a_{n\:out}^{\dag}(-),
$$
$$
a_{n\:in}^{\dag}(-)=-e_{2n}^*a_{n\:out}^{\dag}(+)+e_{1n}^*a_{n\:out}^{\dag}(-).
                                                            \eqno(30)
$$

\section{Matrix elements and probabilities}

We denote  $|0_{n\;in}>$ the in-vacuum state in the cell with 
quantum numbers $n$ and similarly for the out-vacuum. To find 
$<0_{n\;out}|0_{n\;in}>$, we rewrite Bogoliubov transformations 
(21) in the form
$$
a_{n\;out}=B_n^{-1}a_{n\;in}B_n,\quad 
 b^{\dag}_{n\;out}=B_n^{-1}b^{\dag}_{n\;in}B_n,     \eqno(31)
$$
where [3]
$$
B_n=c_{1n}^*+(1-c_{1n}^*)[a_{n\;in}^{\dag}a_{n\;in}+b_{n\;in}^{\dag}b_{n\;in}]-
c_{2n}^*a_{n\;in}^{\dag}b_{n\;in}^{\dag}
$$
$$
-c_{2n}a_{n\;in}b_{n\;in}
+(c_{1n}+c_{1n}^*-2)a_{n\;in}^{\dag}a_{n\;in}b_{n\;in}^{\dag}b_{n\;in}.
\eqno(32)
$$
Eq.(31) imply $<0_{n\;out}|=<0_{n\;in}|B_n$ and hence
 $$
 <0_{n\;out}|0_{n\;in}>=c_{1n}^*.  \eqno(33)
 $$

 We note now that the unitary operator $B$ is defined by (31) and (21) 
 only up to a
 phase factor and we put it equal to unity. It is a natural choice.
  It leads to (33)
 from which the correct vacuum-vacuum amplitude for a constant
 electromagnetic field can be obtained [13].

 Now we can write down matrix elements. We start with the Klein region.
 From second eq. (22) we have
 $$
 b_{n\;in}= 
-c^*_{2n}a_{n\;out}^{\dag}+c_{1n}^*b_{n\;out}.   \eqno(34)  
$$
Using this relation, we find 
$$
b_{n\;out}|0_{n\;in}>=c_{2n}^*c_{1n}^{*-1}a_{n\;out}^{\dag}|0_{n\;in}>.
                                                       \eqno(35)
$$
From here for the  pair creation amplitude we find
 $$
 <0_{n\;out}|a_{n\;out}b_{n\;out}|0_{n\;in}>=c_{2n}^*c_{1n}^{*-1}
 <0_{n\;out}|0_{n\;in}>=c_{2n}^*.                             \eqno(36)
 $$
  The sum of all probabilities in the cell $n$ initially in the vacuum state 
 is 
 $$
 |<0_{n\;out}|0_{n\;in}>|^2+|<0_{n\;out}|a_{n\;out}b_{n\;out}|0_{n\;in}>|^2=1, 
                                                                \eqno(37)
 $$
 see (33), (36) and the first eq. in (19).
 Similarly, when the initial state with quantum numbers $n$ is occupied, 
  we get  for the scattering amplitude
 $$
  <0_{n\;out}|a_{n\;out}a^{\dag}_{n\;in}|0_{n\;in}>=c_{1n}^*{}^{-1}
  <0_{n\;out}|0_{n\;in}>=1.                                       \eqno(38)
  $$
 We see that the 
information on processes in an external field is contained in the solutions of 
wave equation, but it has to be decoded.
In particular, if  the initial state with quantum numbers $n$ is occupied,
we know that the electron cannot penetrate deep into the barrier. It is 
suggested in [14-15] that the accessible region is defined by the condition
$\pi^0(x_3)>0$.

  Now we go to the scattering region, extending its definition to all energies
  outside the Klein region. In this case $c_{2n}=0$,
   $c_{1n}c_{1n}^*=1$ in (18), (19) and (21). Hence
   $$
 {}^+\psi_n=c^*_{1n}\,{}_+\psi_n,\quad  a_{n\;out} =a_{n\;in}c_{1n} \eqno(39)
 $$
 and similarly for the other quantities. We first consider the scattering 
 region, where the reflection is complete. Then $\pi_3(+)$ is imaginary and
 $|c'_{1n}|=|c'_{2n}|$ due to current conservation. The solutions ${}_{\pm}f_n$
 in (13) must be discarded as containing exponentially growing terms when
 $x_3\to\infty$. Instead of two solutions ${}^{\pm}f_n$ we are left with 
 only one as there is only one boundary condition when $x_3\to\infty$, see
 eq. (6).
 The reflection amplitude is
$$
<0_{n\;out}|a_{n\;out}(-)a_{n\;in}^{\dag}(+)|0_{n\;in}>=
c_{1n}<0_{n\;out}|0_{n\;in}>=1,                                      \eqno(40)
$$
see the second eq. in (39), eq. (33) and the condition $c_{1n}c_{1n}^*=1$.

The final state ${}^+\psi_n$ in (39) differs from the initial state
 ${}_+\psi_n$ only by the phase factor
given by the (renormalized) value of $<0_{n\;out}|0_{n\;in}>$, see (33), i.e.
by  $<0_{n\;out}|0_{n\;in}>^{ren}\equiv e^{i\phi_n}$[13].
 In principle
this factor can be observed in the interference pattern of the incident and
reflected waves. Then it will be a way to find $\phi_n $ experimentally.

We note here that the considered region of complete reflection can be reached
from the Klein region by raising the value of $p^0$. At the top of the
Klein region $c_{2n}$ is very small and the reflected wave resembles the one
at the bottom of complete reflection region; the corresponding $c_{1n}$ just
above and below the boundary between two regions must
be almost equal.

By raising $p^0$  still further, we enter into the scattering region where
$\pi^0(\pm)>m_{\perp}$. For the initial state ${}_+\psi_n(x_3|+)$, using (27), 
we find for the reflection and transmission
amplitudes
$$
<0_{n\;out}|a_{n\;out}(-)a_{n\;in}^{\dag}(+)|0_{n\;in}>=
e_{2n}<0_{n\;out}|0_{n\;in}>,                                      \eqno(41)
$$
$$
<0_{n\;out}|a_{n\;out}(+)a_{n\;in}^{\dag}(+)|0_{n\;in}>
=e_{1n}<0_{n\;out}|0_{n\;in}>.                                     \eqno(42)
$$

Alternatively we may say that our solutions ${}^+\psi_n(x_3|\pm)$ are 
relative ones and the absolute
solutions are obtained from them by multiplying by  $e^{i\phi_n}$.
 Then the factor
$<0_{n\;out}|0_{n\;in}>$  in the r.h. sides of (41) and (42) disappears.

The propagator method gives the same results. In the Klein region we start from
${}_+\psi_n$. The Feynman propagator evolves this state to the relative function
 $c_{1n}^{*-1}{}^+\psi_n$ [4]. So the absolute final function is ${}^+\psi_n$.
 In scattering region the relative final wave function is the same as
  the initial one and the absolute final function is $e^{i\phi_n}{}_+\psi_n$.
 It is a happy occasion when 
$ c_{1n}^{*ren}$ gives $e^{i\phi_n} \equiv<0_{n\;out}|0_{n\;in}>^{ren}$.
  In general $e^{i\phi_n}$ have to be found by other means,  see [13].
In quantum mechanics for the state ${}_+\psi_n(x_3|+)$ the amplitudes of
  reflection and transmission are
obtained directly from the first equation in (24), which says:
the reflection (transmission) amplitude is $e_{2n}$ ($e_{1n}$).
In the considered scattering region $e^{i\phi_n}$ is a phase factor.
 We see that vacuum virtual
processes lead to the appearance of an additional phase shift in the reflected
and transmitted  waves. 

In conclusion of this section, we note that in the scattering region instead
 of (32) and (31) we have (again with natural choice of phase factor)
 $$
 A_n=1+(e_{1n}-1)a_{n\;in}^{\dag}(+)a_{n\;in}(+)+
 (e^*_{1n}-1)a_{n\;in}^{\dag}(-)a_{n\;in}(-)+e_{2n}
 a_{n\;in}^{\dag}(-)a_{n\;in}(+)-
 $$
 $$
 e^*_{2n}a_{n\;in}^{\dag}(+)a_{n\;in}(-)+
 [2-(e_{1n}+e^*_{1n})]a_{n\;in}^{\dag}(+)a_{n\;in}(+)
 a_{n\;in}^{\dag}(-)a_{n\;in}(-),                           \eqno(43)
 $$
 $$
 A^{-1}_na_{n\;in}(\pm)A_n=a_{n\;out}(\pm),\quad
  <0_{n\;in}|=<0_{n\;in}|A_n,                                  \eqno(44)
$$
see Appendix 5 in [16].
  
\section{Normalization of wave functions}

For our stationary states the (transition) current conservation
gives
$$
i(p'{}^0-p^0)j^0(f_{n'},f_{n})=-\frac{\partial}{\partial x_3}j_3(f_{n'},f_{n}).
                                                                     \eqno(45)
$$
So 
$$
\int_{-L_d}^{L_u}dx_3j^0(f_{n'},f_{n})=
\frac i{p'{}^0-p^0}[j_3(f_{n'},f_{n}|L_u)
-j_3(f_{n'},f_{n}|-L_d)].              \eqno(46)
$$
First we consider $f_{n'}={}_+f_{n'}$, $f_{n}={}_+f_{n}$  where
$n'=(p'^0,p_1,p_2,r)$, $n=(p^0,p_1,p_2,r)$, $r=1,2$. For $x_3\to-\infty$
we can easily calculate $j_3({}_+f_{n'},{}_+f_{n}|-L_d)$ because ${}_+f_{n'}$,
${}_+f_{n}$ are known in this limit, see eq (6) and (7). The result is:
$$
-j_3({}_+f_{n'},{}_+f_{n}|-L_d)\to-e^{-i[\pi_3(-)-\pi_3'(-)]L_d}=
-e^{-iA_d(p^0-p'^0)},\quad A_d=
\frac{\pi^0(-)+\pi'^0(-)}{\pi_3(-)+\pi'_3(-)}L_d.        \eqno(47)
$$
The arrow means that we put $\pi_3'(-)=\pi_3(-)$, $\pi'^-(-)=\pi^-(-)$
everywhere except in the exponent. This is permissible because when 
$L_d\to\infty$, the effectively nonzero result will be achieved only 
when $\pi'$ will be close to $\pi$. In the last equation in (47) we have used
the relation
$$
\pi_3^2-\pi'{}_3^2=(\pi^0)^2-(\pi'{}^0)^2 =(\pi^0+\pi'{}^0)(p^0-p'{}^0),
                                                                 \eqno(48)
$$
see eqs. (1) and (2).

The contribution from $j_3({}_+f_{n'},{}_+f_{n}|L_u)$ is treated in the 
same manner. Using the first equation in (13), we have
$$
j_3({}_+f_{n'},{}_+f_{n})=c'^*_{1n'}c'_{1n}j_3({}^+f_{n'},{}^+f_{n}) +
c'^*_{2n'}c'_{2n}j_3({}^-f_{n'},{}^-f_{n})+
c'^*_{1n'}c'_{2n}j_3({}^+f_{n'},{}^-f_{n}) +
$$
$$
c'^*_{2n'}c'_{1n}j_3({}^-f_{n'},{}^+f_{n})\to 
|c'_{1n}|^2j_3({}^+f_{n'},{}^+f_{n}) +
|c'_{2n}|^2j_3({}^-f_{n'},{}^-f_{n}).   \eqno(49)
$$
In the Klein region this can be written as follows
$$
j_3({}_+f_{n'},{}_+f_{n}|L_u)\to-|c'_{1n}|^2e^{i[\pi_3(+)-\pi_3'(+)]L_u}
 +|c'_{2n}|^2e^{-i[\pi_3(+)-\pi_3'(+)]L_u}=
 $$
 $$
 -|c'_{1n}|^2e^{-iA_u(p^0-p'{}^0)}
 +|c'_{2n}|^2e^{iA_u(p^0-p'{}^0)},\quad A_u=
 \frac{-\pi^0(+)-\pi'^0(+)}{\pi_3(+)+\pi'_3(+)}L_u.         \eqno(50)
 $$
Noting that $A_d$ and $A_u$  are of the same sign, we fix the relation 
between them: $A_d=A_u$. Using also $|c_{2n}'|^2=|c_{2n}'|^2+1$, we finally 
 obtain from (46)
 $$
 \int_{-\infty}^{\infty}dx_3j^0({}_+f_{n'},{}_+f_{n'})=2\pi|c_{2n}'|^2\delta
 (p^0-p'{}^0)                                                 \eqno(51)
 $$
 in the Klein region where $c_{2n}'$ is the amplitude of the largest wave.
 Thus, choosing ${\cal N}_n=|c_{2n}'|^{-1}$, we normalize $\psi-$functions in
 (17) on $2\pi\delta(p^0-p'{}^0)$.

In the scattering region the amplitude of the largest wave is  $c_{1n}'$ and
we must substitute $c_{2n}'\to c_{1n}'$  in the r.h.s. of (51). This can be
seen as follows. The relation (47) remains unchanged. Instead of (50) we
have
$$
j_3({}_+f_{n'},{}_+f_{n}|L_u)\to|c'_{1n}|^2e^{-iA_u(p^0-p'{}^0)}
 -|c'_{2n}|^2e^{iA_u(p^0-p'{}^0)}.      \eqno(52)
 $$
 Besides, now $A_u=-|A_u|$ and $|c_{1n}'|^2=|c_{2n}'|^2+1$. So the condition
 $A_d=|A_u|$ leads to the stated result. We note here that the condition
 $$
 \frac{\pi^0(-)+\pi'^0(-)}{\pi_3(-)+\pi'_3(-)}L_d=
 \frac{\pi^0(+)+\pi'^0(+)}{\pi_3(+)+\pi'_3(+)}L_u         \eqno(53)
 $$
 is a generalization to the scattering
 region of the free field condition $L_d=L_u$.
In the case of a complete reflection $|c'_{1n}|=|c'_{2n}|$ and (51) remains 
valid.

Going back to the Klein region we can  show that ${}^+\psi_{n'}$  and
 ${}^-\psi_n$, see (17), are orthogonal. 
  As ${}^+\psi_n\propto{}^+f_n$ and ${}^-\psi_n\propto{}_+f_n$ ,
  we write
 $$
 j_3({}^+f_{n'},{}_+f_{n})=-c_{1n'}'j_3({}_+f_{n'},{}_+f_{n})
 -c'^*_{2n'}j_3({}_-f_{n'},{}_+f_{n}),                          \eqno(54)
 $$
 where use has been made of the first equation in (14) with lower sign.
  Arguing as above,
 we have
 $$
 j_3({}^+f_{n'},{}_+f_{n}|-L_d)\to-c_{1n}'e^{i[\pi_3(-)-\pi_3'(-)](-L_d)}=
-c_{1n}'e^{-iA_d(p^0-p'{}^0)}.                                    \eqno(55)
$$
Similarly
$$
 j_3({}^+f_{n'},{}_+f_{n}|L_u)\to
-c_{1n}'e^{-iA_u(p^0-p'{}^0)}.                                   \eqno(56)
$$
For $A_d=A_u$ the difference of the last two expressions is zero. So 
${}^+\psi_{n'}$  and ${}^-\psi_n$ are orthogonal. The same is true for
${}_+\psi_{n'}$  and ${}_-\psi_n$.

In the same manner it easy to show that in the scattering region 
${}^+\psi_{n'}(x_3|+)$ and ${}^+\psi_{n}(x_3|-)$ are orthogonal; the 
same is true for ${}_+\psi_{n'}(x_3|+)$ and ${}_+\psi_{n}(x_3|-)$.

Thus, the orthogonal pairs of functions in the Klein region are 
(${}_+f_{n'},{}^+f_{n}$) and (${}_-f_{n'},{}^-f_n$), and in the
 scattering region
(${}_+f_{n'},{}^-f_n$)  and (${}_-f_{n'},{}^+f_n$). This assertion must be
true for particles of any spin.

For the step potential, considered in Sec.6, the results of this Sec.
were checked by a straightforward calculation of the l.h.s. of (46) with
$j^0(f_{n'},f_n)=f^*_{n'}f_n$ and $L_d, L_u$ equal to infinity.

\section{Solvable potential}

For the potential 
$$
A^0(x_3)=-a\tanh(kx_3) \eqno(57)
$$
the solution ${}_+Q_p(x_3,\lambda)$ of the squared Dirac equation has the 
form
$$
{}_+Q_p(x_3,\lambda)=(-z)^{i\mu}(1-z)^{i\lambda}F(\alpha,\beta;\gamma;z),
                                         \eqno(58)
$$
where
$$
-z=e^{2kx_3},\quad \lambda=\frac{ea}k,\quad\pi^0(\pm)=p^0\pm ea,\quad
2k\nu=\pi_3(+),\quad 2k\mu=\pi_3(-),
$$
$$
\alpha=i\mu+i\nu+i\lambda,\quad \beta=i\mu-i\nu+i\lambda,\quad \gamma=1+2i\mu.
                                                                  \eqno(59)
$$
We assume at first that $\pi_3(\pm)$ are real.
The solution ${}_-Q_p(x_3,\lambda)$ can be obtained from (58) either by
substitution $\mu\to-\mu$ or by complex conjugation and substitutions 
$e\to-e$, $p^0\to-p^0$. 

The solution ${}^+Q_p(x_3,\lambda)$ is obtainable from ${}_+Q_p(x_3,\lambda)$
 by complex conjugation and substitutions 
$e\to-e$, $x_3\to-x_3$ (these operations do not change eq. (5) with
$F(-\varphi)=-F(\varphi)$):
$$
{}^+Q_p(x_3,\lambda)=(-z^{-1})^{-i\nu}(1-z^{-1})^{i\lambda}
F(-i\mu-i\nu+i\lambda,i\mu-i\nu+i\lambda;1-2i\nu;z^{-1}).        \eqno(60)
$$
${}^-Q_p(x_3,\lambda)$ is obtainable from (60) by substitution $\nu\to-\nu$.
Now, eq. (11) takes the form
$$
{}_+Q_p(x_3,\lambda)=\frac{\Gamma(1+2i\mu)\Gamma(2i\nu)}
{\Gamma(i\mu+i\nu+i\lambda)
\Gamma(1+i\mu+i\nu-i\lambda)}                    
 {}^+Q_p(x_3,\lambda)
 $$
 $$
 +\frac{\Gamma(1+2i\mu)\Gamma(-2i\nu)}
 {\Gamma(i\mu-i\nu+i\lambda)
\Gamma(1+i\mu-i\nu-i\lambda)}{}^-Q_p(x_2,\lambda).                    \eqno(61)
$$
From here by complex conjugation and substitutions $e\to-e,$ and $x_3\to- x_3$
we find
$$
{}^+Q_p(x_3,\lambda)=\frac{\Gamma(-2i\mu)\Gamma(1-2i\nu)}
{\Gamma(-i\mu-i\nu+i\lambda)
\Gamma(1-i\mu-i\nu-i\lambda)}                    
 {}_+Q_p(x_2,\lambda)
 $$
 $$
 +\frac{\Gamma(1-2i\nu)\Gamma(-2i\mu)}
 {\Gamma(i\mu-i\nu+i\lambda)
\Gamma(1+i\mu-i\nu-i\lambda)}{}_-Q_p(x_2,\lambda).                 \eqno(61')
$$
Neglecting $\lambda=\frac{ea}{kc\hbar}$ we get the nonrelativistic limit, see
Problem 3 in \S 25 in [17] where the potential $U$ differs from our
$eA^0$  only by an additive constant.

Now we can rewrite eq. (15) in the form
$$
c'_{1n}=\left(\frac{\pm\pi^{-}(+)}{\pi^-(-)}
\frac{\pi_3(+)}{\pi_3(-)}
\right)^{1/2}a(\lambda),\quad
c'_{2n}=\left(\frac{\pm\pi^{+}(+)}{\pi^-(-)}
\frac{\pi_3(+)}{\pi_3(-)}\right)
^{1/2}b(\lambda).
                                               \eqno(62)
$$
The sign + ($-$) in front of $\pi^-(+)$ and $\pi^+(+)$ refers to the 
scattering (Klein) region.

We note now that in the Klein region
$$
\pi^0(-)-\pi^0(+)=-2ea=2|ea|     \eqno(63)
$$
is the total energy of the $e^+e^-$ pair. Hence $2|ea|>\pi_3(-)+\pi_3(+)$,
i.e. $-\lambda-\mu-\nu>0$ and all the more so $-\lambda\pm\mu\mp\nu>0$.

In the scattering region $\pi^+(+)/\pi^-(-)>0$ and from here we shell obtain
$$
\mu-\nu+\lambda>0.         \eqno(64)
$$
 To this end we use the relation [4]
 $$
 \frac{\pi^+(+)}{\pi^-(-)}=\frac{\pi^+(-)}{\pi^-(+)}=
 \frac{\mu-\nu-\lambda}{\mu-\nu+\lambda}=
 \frac {\pi_3(-)-\pi_3(+)-2ea}{\pi_3(-)-\pi_3(+)+2ea}.          \eqno(65)
 $$
 In the scattering region $\pi^0(-)=\pi^0(+)+2|ea|$, see (63), i.e.
   $\pi^0(-)>\pi^0(+)$. Hence, $\pi_3(-)>\pi_3(+)$ and
 all the more so  $\pi_3(-)-\pi_3(+)+2|ea|>0$. The numerator in the r,h.s.
 of (65) is positive, so must be the denominator because the l.h.s. of the
 expression is positive.

 We note also another useful relation, obtainable from (65) by substitution
 $\pi_3(+)\to-\pi_3(+)$
 $$
  \frac{\pi^-(+)}{\pi^-(-)}=\frac{\pi^+(-)}{\pi^+(+)}=
 \frac{\mu+\nu-\lambda}{\mu+\nu+\lambda}.          \eqno(66)
 $$

With the help of these relations we find more symmetric expressions for
$c'_{1n}$, $c'_{2n}$
$$
c'_{1n}=2\sqrt{\frac{\mu\nu}{\pm(\mu+\nu+\lambda)(\mu+\nu-\lambda)}}
\frac{\Gamma(2i\mu)\Gamma(2i\nu)}{\Gamma(i\mu+i\nu+i\lambda)
\Gamma(i\mu+i\nu-i\lambda)},
$$
$$
c'_{2n}=2\sqrt{\frac{\mu\nu}{\pm(\mu-\nu+\lambda)(\mu-\nu-\lambda)}}
\frac{\Gamma(2i\mu)\Gamma(-2i\nu)}{\Gamma(i\mu-i\nu+i\lambda)
\Gamma(i\mu-i\nu-i\lambda)}.                    \eqno(67)
$$
Here and below the upper sign  refers to the scattering region,
the lower one to the Klein region. 
We see that
$$
c'_{1n}(\mu,\nu,\lambda)=c'_{1n}(\nu,\mu,\lambda),\quad
c'{}_{2n}(\mu,\nu,\lambda)=c'{}^*_{2n}(\nu,\mu,\lambda).       \eqno(68)
$$
It is still assumed that $\mu, \nu$ and $ \lambda$ are real.

When we approach  the complete scattering region from the Klein region, 
$\pi_3(+)\to0$, and
$c'_{1n}$, $c'_{2n}$ become unlimited. Then $c_{2n}\to0$ and $c_{1n}$ becomes
a phase factor, see eq.(19).

When  $k\to0$, $a\to\infty$, $ak=E$ in eq. (57), we get the potential of a
constant electric field. This case was studied earlier, see [4], [11], [13].
When $k\to\infty$, we have a step potential. This case is instructive because
of its simplicity. In particular, we can verify the orthogonality and
normalization in the set of solutions by a straightforward calculations.

\section{Step potential}
 From (62), (61) and (59) we find when $k\to\infty$
 $$
c'_{1n}=\frac12\left(\frac{\pm\pi^{-}(+)}{\pi^-(-)}
\frac{\pi_3(+)}{\pi_3(-)}
\right)^{1/2}\left(1+\frac{\pi_3(-)+2ea}{\pi_3(+)}\right),
$$
 $$
c'_{2n}=\frac12\left(\frac{\pm\pi^{+}(+)}{\pi^-(-)}
\frac{\pi_3(+)}{\pi_3(-)}
\right)^{1/2}\left(1-\frac{\pi_3(-)+2ea}{\pi_3(+)}\right).  \eqno(69)
$$

Due to relations (65) and (66) there are several other useful forms for 
$c'_{1n}$ and $c'_{2n}$.
We note these
$$
c'_{1n}=\pm\frac12\left(\frac{\pm\{[\pi_3(-)+\pi_3(+)]^2-(2ea)^2\}}
{\pi_3(-)\pi_3(+)}\right)^{1/2},
$$
$$
c'_{2n}=\mp\frac12\left(\frac{\pm\{[\pi_3(-)-\pi_3(+)]^2-(2ea)^2\}}
{\pi_3(-)\pi_3(+)}\right)^{1/2},                                \eqno(70)
$$
and these
 $$
c'_{1n}=\pm\frac12\left(\frac{\pi^-(-)}{\pm\pi^{-}(+)}
\frac{\pi_3(-)}{\pi_3(+)}
\right)^{1/2}\left(1+\frac{\pi_3(+)-2ea}{\pi_3(-)}\right),
$$
 $$
c'_{2n}=-\frac12\left(\frac{\pi^+(-)}{\pm\pi^{-}(+)}
\frac{\pi_3(-)}{\pi_3(+)}
\right)^{1/2}\left(1-\frac{\pi_3(+)-2ea}{\pi_3(-)}\right). \eqno(71)
$$

 From (70) and (19)  we find (in both regions)
 $$
 c_{1n}=-\frac{c'_{1n}}{c'{}^*_{2n}}=
 \left(\frac{(2ea)^2-[\pi_3(-)+\pi_3(+)]^2}{(2ea)^2-[\pi_3(-)-\pi_3(+)]^2
 }\right)^{1/2}.          \eqno(72)
 $$
 We see that $c_{1n}\to1$ when $\pi_3(+)\to0$ and probably in this limit 
 $e^{i\phi_n}=1$. 

As mentioned at the end of Sec. 3 the results of that section were verified,
using step potential. The expressions (69) and (71) turns out to be most
handy, cf. examples of much less cumbersome calculations for a scalar particle
in the first reference in [9].

\section{Conclusion}
 In general the $S-$matrix approach gives the same results as the propagator
 method. Yet, the former approach gives naturally the expression for 
 $<0_{n\;out}|0_{n\;in}>$. Its renormalized value provides the factor
  distinguishing absolute and relative wave functions. The contribution 
  from (virtual) pair production to any given final wave function must 
  be observable in principle. If the theory is correct, it suggests that
  particle clock ticking depends upon the field.
 
\section*{Acknowledgments}
I am grateful to V.I.Ritus for encouraging discussions and useful suggestions.
This work was supported in part by the Russian Foundation for Basic Research
(projects no. 00-15-96566 and 02-02-16944)

 \section*{References}
\begin{enumerate}

\item  N.B.Narozhny and A.I.Nikishov, Yad. Fiz. {\bf 11}, 1072 (1970). \\
\item A.I.Nikishov, Nucl. Phys. {\bf B 21}, 346 (1970). \\
\item A.I.Nikishov, Tr. Fiz. Inst. Akad. Nauk SSSR  {\bf 168}, 157 (1985); \\
  in {\sl Issues in Intensive-Field Quantum Electrodynamics}, Ed. by
  V.L.Ginzburg (Nova Science, Commack, 1987).   \\
\item
A.I.Nikishov, Tr. Fiz. Inst. Akad. Nauk SSSR {\bf 111}, 152 (1979); J. Sov.
 Laser Res. {\bf 6},\\
 619 (1985). \\
\item A.A.Grib, S.G.Mamaev, and V.M.Mostepanenco,{\sl Vacuum Quantum Effects
  in Strong Fields} (Energoatomizdat, Moscow, 1988).\\
\item  A.Hansen, F.Ravndal, Physica Scripta, {\bf 23}, 1036 (1981).      \\
\item  W.Greiner, B.M\"uller, J.Rafelski, {\sl Quantum Electrodynamics of Strong
    Field}, Springer-Verlag (1985). \\
\item  A.Calogeracos, N.Dombey, Contemp. Phys. {\bf 40}, 313 (1999).    \\
\item  A.I.Nikishov, hep-th/0111137; A.I.Nikishov, {\sl Problems of atomic
science and technology}, Special issue dedicated to the 90-birthday 
anniversary of A.I.Akhieser, Kharkov, Ukraine, p.103 (2001).\\
\item A.I.Nikishov, hep-th/0202024.\\
\item A.I.Nikishov, hep-th/0211088.\\
\item A.I.Nikishov, Teor. Mat. Fiz. {\bf 98}, 60 (1994).\\
\item A.I.Nikishov, hep-th/0207085.\\
\item A.I.Nikishov, Zh.\' Exsp. Teor. Fiz. {\bf91}, 1565 (1986)
[Sov. Phys. JETP  {\bf 64}, 922 (1986)]. \\
\item A.I.Nikishov, Yad. Fiz. {\bf 46}, 163 (1987) [Sov. J. Nucl. Phys.
{\bf46}, 101 (1987)].\\ 
\item F.A.Kaempfer, {\sl Concepts in quantum mechanics}, 
Academic press, New York (1965). \\
\item L.D.Landau, E.M.Lifshits, {\sl Quantum mechanics}, Pergamon Press, 1977.
 \end{enumerate}
\end{document}